\documentclass[aps,prl,twocolumn,tightenlines,nofootinbib,superscriptaddress]{revtex4}
\usepackage{epsfig}
\usepackage{graphicx}
\usepackage{psfrag}
\usepackage{amsmath,amssymb,psfrag,slashed,graphicx}
\usepackage{colordvi}
\usepackage{color}

\usepackage{slashed}
\usepackage{color}
\usepackage[normalem]{ulem}

\begin{document}

\title{Transverse spin sum rule of the proton}

\author{Xiangdong Ji}
\email{xji@umd.edu}
\affiliation{Center for Nuclear Femtography, SURA, 1201 New York Ave. NW, Washington, DC 20005, USA}
\affiliation{Department of Physics, University of Maryland, College Park, MD 20742, USA}

\author{Feng Yuan}
\email{fyuan@lbl.gov}
\affiliation{Nuclear Science Division, Lawrence Berkeley National Laboratory, Berkeley, CA 94720, USA}

\date{\today}

\begin{abstract}

The transversely-polarized state of a proton with arbitrary momentum
is not an eigenstate of transverse angular momentum operator. The latter
does not commute with the QCD Hamiltonian. However, the expectation value of the
transverse angular momentum in the state is well-defined and grows
proportionally to the energy of the particle. The transverse
spin content of the proton is analyzed in terms of the QCD angular momentum structure.
In particular, we reconfirm that the generalized parton distributions $H+E$ provide the leading-twist
transverse angular momentum densities of quarks and gluons in
the infinite momentum frame.

\end{abstract}

\maketitle

\section{Introduction}
The spin structure of the proton has been an important subject in
hadronic physics for more than 30 years.
Much of the discussion so far has been focused on the proton
helicity, the projection of spin or total angular momentum (AM) along
the direction of motion, in a longitudinally-polarized state.
Sum rules have been derived for the proton helicity~\cite{Jaffe:1989jz,Ji:1996ek},
and many experiments have been carried out to measure various contributions.
Reviews of the proton spin physics can be found in Refs.~\cite{Filippone:2001ux,Bass:2004xa,Aidala:2012mv,Leader:2013jra,Deur:2018roz,Ji:2020ena}.

There is much less discussion about transverse spin structure
of the proton, except for various spin phenomena related to
transverse polarization, such as $g_2(x)$ structure function~\cite{Jaffe:1990qh}, transversity distribution~\cite{Artru:1989zv,Cortes:1991ja,Jaffe:1991kp,Jaffe:1991ra},
and single spin asymmetries~\cite{Sivers:1989cc,Collins:1992kk}.
The existing results on the transverse spin content of the proton in the literature~\cite{Bakker:2004ib,Burkardt:2002hr,Burkardt:2005hp,Leader:2011cr,Ji:2012sj,Ji:2012vj,Hatta:2012jm,Harindranath:2013goa}
are controversial.

In this paper, we derive a number of results about the transverse
AM in a transversely-polarized proton. In one of our previous papers on this~\cite{Ji:2012vj},
we started from the Pauli-Lubanski (PL) spin because its projection along the transverse
direction defines the transverse polarization. However, since the operator involves
Lorentz boost which has dubious physical meaning in a bound state, we focus
here on the transverse AM operator.
We find that the expectation value of the latter is well-defined and grows
proportionally to the energy of the particle. Thus, we analyze the transverse
spin content in terms of the QCD transverse AM, and derive a partonic sum rule in the infinite momentum frame (IMF)~\cite{Ji:2012sj,Ji:2012vj}.
We also show that
canonical AM decomposition in the light-cone gauge $A^+=0$
and IMF gives the same result.

The key point in our derivation is to separate the intrinsic transverse spin
contribution from the center-of-mass motion contribution to
the transverse AM of the nucleon in a moving frame. Because of
Lorentz invariance, these two components can be distinguished by their different
behaviors under the boost along the momentum direction of the nucleon.

We emphasize that it is important to distinguish three related concepts:
transverse polarization, transverse AM and transverse spin.
Transverse polarization is defined through the PL spin vector, $W^\mu$.
Transverse AM is a part of the PL spin vector which also contains the boost operator.
Finally, we refer the transverse spin as $\hbar/2$, the ratio of the intrinsic
part of the transverse AM over Lorentz boost factor $\gamma$.

The rest of this paper is organized as follows. In Sec.~II, we review the basic
formalism of the nucleon spin constructed from the projection of the PL
vector and its application to the transversely-polarized proton. We derive, to our knowledge,
an important new result in Eq. (15). In Sec.~III, We discuss the QCD source
of AM and derive the transverse spin sum rule for proton at any momentum.
A partonic sum rule in terms of generalized parton distributions~\cite{Ji:1996ek,Ji:1996nm}
will be discussed in Sec.~IV. Finally, we summarize our paper in Sec. V.

\section{Transverse Polarization and angular momentum expectation}

To discuss the spin structure of the proton, we start from
the spin as a emergent concept from the symmetry of space and
time~\cite{Tung:1985na}. The relativistic spin operator $W^\mu$ is the PL four-vector,
\begin{align}
 W^\mu &= -\frac{1}{2}\epsilon^{\mu\alpha\lambda\sigma} J_{\alpha\lambda} P_\sigma/M  \\
   &= \gamma(\vec{J}\cdot \vec{\beta}, ~~\vec{J} + \vec{K}\times \vec{\beta}) \ ,
\end{align}
where the fully anti-symmetric Levi-Civita symbol $\epsilon^{0123}=1$, $\vec{K}$ is the Lorentz boost operator,
$\vec{J}$ is the AM operator and $M$ is the particle mass. $K^i$ and $J^i$ with $i=1,2,3$ are related to the Lorentz generators
by $K^i=J^{0i}$ and $J^i=\epsilon^{0jkl} J^{kl}/2$. In the second line of the above
equation, we have replaced the four-momentum operator $P_\sigma$ by
its eigenvalues, the velocity $\vec{\beta} = \vec{v}/c$ and boost factor $\gamma =
(1-\beta^2)^{-1/2}$, which specify a Lorentz frame.
The frame-independent concept of spin is related to a conserved
scalar operator $W^\mu W_\mu$ which has eigenvalue $-s(s+1)\hbar^2$
independent of the particle's momentum, where $s=0,1/2,1,...$ is the spin
quantum number~\cite{Tung:1985na}.
Thus spin is not only related to AM but also to boost.
We will focus on developing
a spin picture in a general Lorentz frame in terms of the AM operator alone,
because the physics of the boost operator is less clear in the bound state structure.

In the rest frame, the proton spin state $|\vec{\beta}=0,\vec{s}\rangle$
can be defined with the AM quantized along $\vec{s}$,
\begin{equation}
  \vec{s}\cdot \vec{J} \left|\vec{\beta}=0,\vec{s}\right\rangle = ({\hbar}/{2}) \left|\vec{\beta}=0,\vec{s}\right\rangle \ .
\end{equation}
Boosting the above to an arbitrary Lorentz frame,
one has
\begin{equation}
  (-W^\mu S_\mu) |PS\rangle = ({\hbar}/{2})|PS\rangle\ ,
\label{eq:spineigen}
\end{equation}
where $|PS\rangle$ have definite momentum $P^\mu=M\gamma(1,\vec{\beta})=(E,\vec{P})$ and polarization four-vector
\begin{equation}
S^\mu =(\gamma\vec{s}\cdot \vec{\beta},~~\vec{s} +(\gamma-1) \vec{s}\cdot \hat \beta\hat \beta) \ ,
\end{equation}
with $\hat{\beta}=\vec{\beta}/|\vec{\beta}|$, $S^\mu S_\mu=-1$ and $P^\mu S_\mu=0$. The normalization
of the state is covariant, $\langle PS|PS\rangle = 2E(2\pi)^3\delta^3(0)$.

In terms of the polarization vector, the generalized spin projection operator is
\begin{equation}
  S_p=-W^\mu S_\mu = \gamma \vec{J}\cdot \vec{s} - (\gamma-1)\vec{s}\cdot \vec{\beta} \vec{J}\cdot \vec{\beta}
     - \gamma \vec{s}\cdot (\vec{K}\times \vec{\beta}) \ .
\end{equation}
Although a general polarization state $|PS\rangle$ is an
eigenstate of $S_p$, the operator contains not only the AM operator,
but also the boost. As such, $S_p$ in general
is not a good operator for studying the spin structure of the proton.

Without loss of generality, one can assume the proton momentum to be
$\vec{P} = (0,0,P^z)$.
For transverse polarization along the $x$-direction,  we have
\begin{equation}
            \vec{s}_x = (1,0,0) \ .
\end{equation}
And the covariant polarization four-vector is
\begin{equation}
             S^\mu_x = (0, \vec{s}_x) \ ,
\end{equation}
which does not grow with the particle's momentum.
Therefore, the transverse polarization vector
appears to be sub-leading or twist-three. In the parton language, the
leading twist has simple partonic picture, whereas the
subleading twist is related to parton correlations.
However, we emphasize that this counting is
only useful to a certain extent for spin physics,
and is not true in all cases.

In the rest frame, one has,
\begin{eqnarray}
   \langle \beta=0,\vec{s}_x|J^x|\beta=0,\vec{s}_x\rangle &=& \langle J^x\rangle=\hbar/2\ ,  \\
    \langle \beta=0,\vec{s}_x|K^y|\beta=0,\vec{s}_x\rangle &=& \langle K^y\rangle = 0 \ ,
\end{eqnarray}
where, for simplicity, here and in the rest of the section, we have omitted the normalization of the state (divided by
$\langle PS|PS\rangle$ on the left hand side).
It must be pointed out that the above results
are only true for the intrinsic spin part. In particular, the second
equation comes from the fact that the spin part of $K^y$ couples
the upper and lower components of the Dirac spinor which has only the upper
one in the rest frame.

Since $\vec{K}$ and $\vec{J}$ transform
under Lorentz tranformation as $(1,0)+(0,1)$,
we have
\begin{eqnarray}
     \langle J^{x}\rangle' &=& \gamma  ( \langle J^{x}\rangle - \beta  \langle K^{y}\rangle  ) \ ,\\
     \langle K^{y}\rangle' &=& \gamma  ( \langle K^{y}\rangle + \beta  \langle J^{x}\rangle  ) \ .
\end{eqnarray}
From this, we can deduce that in a moving proton:
\begin{eqnarray}
\langle PS_x |J^x|PS_x \rangle &=&\gamma (\hbar/2)\ ,  \\
\langle PS_x |K^y|PS_x\rangle &=& \gamma\beta(\hbar /2) \ .
\end{eqnarray}
Again, we emphasize this is only true for the intrinsic part
of the spin. In fact, it is easy to verify that
the above are consistent with
Eq.~(\ref{eq:spineigen}) that
a transversely-polarized proton
is an eigenstate of $S_p =  \gamma(J^x-\beta K^y)$.

Therefore to understand the transverse spin structure
of the proton, a natural starting
point is
\begin{equation}
        \langle PS_\perp |J^\perp |PS_\perp \rangle  = \gamma ({\hbar}/{2})  \ ,
\end{equation}
where $\perp$ can be any direction transverse to the direction of motion ($z$).
The above shows that the intrinsic transverse AM
grows as the energy of the particle, a fact less
appreciated in the literature so far. This
indicates that there shall be a simple
partonic interpretation for the transverse spin~\cite{Ji:2012vj}.

\section{QCD Angular Momentum and Transverse Spin Sum Rule}

To understand the proton spin,
we can start from QCD AM operator expressed
in terms of individual parts,
\begin{equation}
    \vec{J}_{\rm QCD} =\sum_\alpha \vec{J}_\alpha \ .
\end{equation}
Through the above, one can express the transverse spin
$\hbar/2$ as contributions from different sources.
This has been a standard approach
to explore the origins of the proton spin
in the literature.

From the standard QCD lagrangian, a straightforward calculation yields the canonical AM expression~\cite{Jaffe:1989jz}
\begin{align}
\vec{J}_{\rm QCD} =& \int d^3 \vec{x}\left[\psi^\dagger_f\frac{\vec{\Sigma}}{2} \psi_f
                + \psi^\dagger_f \vec{x} \times (-i\vec{\partial})\psi_f \right.\nonumber \\
                & \left. + \vec{E}_a\times \vec{A}_a + E^i_a(\vec{x}\times \vec{\partial})A^i_a\right] \ ,
\label{eq:amcanonical}
\end{align}
where $\vec{\Sigma}={\rm diag}(\vec{\sigma},\vec{\sigma})$ with $\vec{\sigma}$ being the Pauli matrix, and the contraction of flavor and color indices, as well as the spatial Lorentz index ``$i$'', is implied. The above expression
contains four different terms, each of which has a clear physical
meaning in free-field theory. The first term corresponds to the quark spin,
the second to the quark orbital AM (OAM), the third to the gluon spin,
and the last one to the gluon OAM. Apart from the
first term, the rest are not manifestly gauge-invariant under the general
gauge transformation.  However, the total
is invariant under the gauge transformation up to a surface
term at infinity which can be ignored in physical-state matrix elements.

Using the Belinfante improvement procedure~\cite{Belinfante:1940},
one can obtain a gauge-invariant form~\cite{Ji:1996ek},
\begin{align}
\vec{J}_{\rm QCD}& = \vec{J}_q+\vec{J}_g \nonumber \\& =\int d^3 x \left[ \psi^\dagger_f \frac{\vec{\Sigma} }{2}\psi_f +
\psi^\dagger_f \vec{x} \times (-i\vec{\nabla}-g\vec{A}) \psi_f \right. \nonumber\\
& \left.+  \vec{x}\times(\vec{E}\times\vec{B})\right]\ .
\label{eq:amgi}
\end{align}
All terms are manifestly gauge invariant, with the second term as mechanical or kinetic
OAM, and the third term gluon AM.

To evaluate the quark orbital and gluon contributions, we introduce the
AM density, $M^{\mu\nu\lambda}$, of QCD, from which the AM operator is defined.
It is well-known that the AM density is related to the energy-momentum tensor (EMT)
$T^{\mu\nu}$ through~\cite{Belinfante:1940,Jaffe:1989jz},
\begin{equation}
M^{\mu\nu\lambda}(x)=x^\nu T^{\mu\lambda}-x^\lambda T^{\mu\nu} \,.
\end{equation}
The individual contributions to the EMT, hence AM density,
can be written as the sum of quark and gluon parts,
\begin{align}
T^{\mu\nu}=T^{\mu\nu}_q+T^{\mu\nu}_g \ ,\label{tmu}
\end{align}
where
\begin{align}
T^{\mu\nu}_q&=\frac{1}{2}\left[\bar\psi\gamma^{(\mu}i\overrightarrow{D}^{\nu)}\psi+\bar\psi\gamma^{(\mu}i\overleftarrow{D}^{\nu)}\psi\right]\,,\\
T^{\mu\nu}_g&=\frac{1}{4}F^2g^{\mu\nu}-F^{\mu\alpha}{F^{\nu}}_{\alpha} \ ,\label{tmuqg}
\end{align}
where $T_q$ includes quarks of all flavor.
The expectation values of the AM densities can be derived from the off-forward matrix elements of EMT~\cite{Ji:1996ek},
\begin{align}
& \langle P'S|T^{\mu\nu}_{q/g}(0)|PS\rangle=\bar U(P'S) \left[A_{q/g}(\Delta^2)\gamma^{(\mu}\bar P^{\nu)}\right.
  \nonumber \\
& \quad +B_{q/g}(\Delta^2)\frac{\bar P^{(\mu}i\sigma^{\nu)\alpha}\Delta_\alpha}{2M}
+C_{q/g}(\Delta^2)\frac{\Delta^\mu\Delta^\nu-g^{\mu\nu}\Delta^2}{M} \nonumber \\
&\quad \left.+\bar C_{q/g}(\Delta^2)Mg^{\mu\nu}\right]U(PS) \  , \label{energy}
\end{align}
where $\bar P=(P+P')/2$, $\Delta=P'-P$. $U$ and $\bar{U}$ are Dirac spinors for the nucleon state, and $A$, $B$, $C$ and $\bar C$
are four independent form factors.

Now, we consider the contributions of the quark and gluon AM to the transverse spin
of the proton.
Define,
\begin{equation}
    J^{q,g}_\perp = \langle PS_\perp|{J}_{\perp q,g}|PS_\perp\rangle /(\gamma\langle PS|PS\rangle)\,,
\end{equation}
then the quark and gluon contributions can
be related to the form factors in Eq.~(\ref{energy}).
In calculating the matrix element above
one must take into account the following observation.
The intrinsic part of the spin is what's Pauli-Lubanski vector
ensures through the projection $\epsilon^{\alpha\beta\gamma\delta}J_{\beta\gamma}P_\delta$.
Thus any contribution from the matrix elements of $J_{\beta\gamma}$ that is
proportional to $P_\beta$ and $P_\gamma$ are not intrinsic, and
must be dropped, as these contributions are coming
from the center-of-mass motion of the proton. We notice that $J_{\beta\gamma}$ is constructed from the energy-momentum tensor through $J_{\beta\gamma}\propto \int d^4x \left[x^\beta T^{0\gamma}-x^\gamma T^{0\beta}\right]$. Furthermore, we can simplify Eq.~(23) as,
\begin{eqnarray}
&&\langle P' |T^{\mu\nu}|P\rangle
   = i\frac{\Delta_\rho}{2M}\left[ (A+B)(P^\mu \epsilon^{\rho\nu\alpha\beta}+P^\nu \epsilon^{\rho\mu\alpha\beta})\right. \nonumber \\
  &&          \left.  + 2\epsilon^{0\rho\alpha\beta}(AP^\mu P^\nu + M^2\bar C g^{\mu\nu})/(E+M)\right]S_\alpha P_\beta \nonumber \\
 &&  + \cdots \label{eq:e25}
\end{eqnarray}
Clearly, any term proportional to $P^\nu$ (which in $J_{\beta\gamma}$ becomes a term
proportional $P_\beta$ or $P_\gamma$) does not contribute to the intrinsic properties, and the first term in the second line of the above equation drops out. Now, if we focus on the transverse spin, which is constructed as
\begin{eqnarray}
 \langle J^x \rangle &= \langle J^{yz}\rangle =\langle M^{0yz} \rangle \ . 
\label{eq:amtensor}
\end{eqnarray}
We find that the $g^{\mu\nu}$ term in Eq.~(\ref{eq:e25}) does not contribute either. Therefore, we will only have contribution from the first term and
one finds,
\begin{equation}
      J^{q, g}_\perp = (A_{q,g}+B_{q,g})/2 \,, 
\end{equation}
which is independent of the momentum of the proton and is
exactly the same as the helicity case~\cite{Ji:1996ek}. Thus one has the transverse spin sum rule
\begin{equation}
     J^q_\perp + J^g_\perp = \hbar/2 \ .
\label{eq:transversesumrule}
\end{equation}
The $\bar C$ does not contribute because it does not contribute to
the momentum density which is the source of AM~\footnote{If $M^{+yz}$ is used
to construct the transverse angular momentum, there will be a potential
contribution from $\overline{C}$ form factor~\cite{Leader:2011cr,Hatta:2012jm}.}.
The potential contribution from the center-of-mass
motion has lead to the incorrect results for the transverse
AM in the literature~\cite{Bakker:2004ib,Leader:2011cr}.

One may further decompose the quark contribution into transverse spin
$\Delta\Sigma_\perp/2 $ and orbital ($L_{q\perp}$) ones,
\begin{equation}
     J^q_\perp = \frac{1}{2}\Delta \Sigma_\perp +L_{q\perp} \ .
\end{equation}
It is easy to derive that the former from~\cite{Jaffe:1989jz}
\begin{equation}
    \langle PS_\perp|\overline{\psi}\gamma^\mu\gamma_5\psi|PS_\perp\rangle
     = \Delta\Sigma M S^\mu \ ,
\end{equation}
where $\Delta\Sigma$ is the singlet axial charge of the nucleon, from which we have
\begin{equation}
     \Delta \Sigma_\perp = (M/E)^2\Delta \Sigma =\gamma^{-2}\Delta\Sigma  \ .
\end{equation}
In the rest frame, this is exactly the same as the quark helicity contribution, as it must be.
As the momentum of the proton gets large, the transverse
spin contribution is suppressed by $1/\gamma^2$, although the sum
of the spin and orbital contribution in Eq. (27) is frame independent.
In particular,  in the IMF, the entire quark contribution is from the orbital motion, $L_{q\perp}$.

\section{Partonic sum rule for transverse spin in Infinite Momentum Frame}

Since the transverse AM grows with energy of a
particle, it should have a natural interpretation in terms of
partons in the IMF. This is a actually a subtle subject~\cite{Ji:2012vj}.
Consider the transverse AM in term of EMT in
Eq.(\ref{eq:amtensor}), it is clear that
$T^{0z}$ has a leading partonic interpretation, whereas the other part
$T^{0y}$ does not because $T^{0y}$ involves transverse momentum
which is a subleading (twist-three) property of a parton. However,
the contribution from these two sources are exactly the same
from symmetry reason, namely they both contribute to 1/2 of
the transverse AM. Here we focus on the part of the transverse
angular momentum that has a leading parton density interpretation, leaving
the twist-three contribution for a future discussion.

In the IMF, $P^z\to \infty$, both $z$ and $0$
components are leading. Consider the gauge-invariant form of the EMT,
the quark part is
\begin{equation}
   T^{0z}_q = \frac{1}{2}\overline{\psi}(\gamma^0 iD^z + \gamma^z iD^0)\psi  \ .
\end{equation}
When this operator is boosted to IMF~\cite{Ji:2013dva, Ji:2014gla}, the leading contribution is the same as
\begin{equation}
   T^{++}_q = \overline{\psi}\gamma^+ iD^+\psi \ ,
\end{equation}
which in the light-cone gauge $A^+=0$ becomes
\begin{equation}
T^{++}_q  = \overline{\psi}\gamma^+ i\partial^+ \psi \ .
\end{equation}
This is the total quark momentum density operator. Likewise, considering
the quark part of the canonical orbital angular momentum in Eq. (\ref{eq:amcanonical}) in IMF.
The part which has a partonic interpretation is
$\int y\overline{\psi}\gamma^0(i\partial)^z\psi$. The momentum density
$\overline{\psi}\gamma^0 i\partial^z\psi$ in the light-cone gauge
is exactly the same as $T^{++}_q$. Therefore, in this special
limit of IMF and light-front gauge, both form of QCD AM reduces
to the same simple non-interacting partonic form.

Therefore, it is natural to define a twist-two
partonic density~\cite{Hoodbhoy:1998yb,Ji:2012sj,Ji:2012vj},
\begin{equation}
   J^{2q}_\perp (x) = x\left[q(x)+ E_q(x)\right]/2,
 \end{equation}
where $q(x)$ is the unpolarized quark/antiquark distributions,
and $E_{q}$ is a type of generalized parton distributions (GPDs).
Integrating the above over $x$ gives the 1/2 of the total
transverse AM carried by quarks,
\begin{equation}
   J^q_\perp = \frac{1}{2}\int dx J^{2q}_\perp (x) + ({\rm twist-3})\ .
\end{equation}
Here, $J^{2q}_\perp(x)$ is the transverse AM density carried by quark parton of momentum
$x$ in a transversely polarized nucleon. Since the twist-three part in the above equation
contributes the same total as the twist-two in the first term, it is tempting 
to regard $J^{2q}_\perp(x)$ alone as the total transverse AM density carried by quark partons. 
\begin{equation}
   J^q_\perp = \int dx J^{2q}_\perp (x) \ . 
\end{equation}
However, it is interesting to work out the twist-three contribution and consider 
its experimental measurement.

Similar argument works for the gluons. The momentum density
of the gluon is
\begin{equation}
   T^{0z}_g = F^{0i}F^{zi} = -E^i F^{zi} \ .
\end{equation}
In the $A^z=0$ gauge which becomes the light-cone guage in the IMF,
the above becomes
\begin{equation}
   T^{0z}_g = F^{0i}F^{zi} = E^i \partial^zA^i \ ,
\end{equation}
which corresponds to exactly the momentum density in the canonical
AM expression in Eq. (\ref{eq:amcanonical}). [The gluon transverse spin contribution
(the $\vec{E}\times \vec{A}$ term) also
vanish like $\gamma^{-2}$ in the IMF.]
Therefore, one can define a gluon transverse AM density,
\begin{equation}
   J^{2g}_\perp (x) = x\left[g(x)+ E_g(x)\right]/2,
\end{equation}
whose integral yields the gluon orbital contribution to the transverse
proton spin,
\begin{equation}
   J^{g}_\perp = \frac{1}{2}\int dx J^{2g}_\perp (x) +({\rm twist-3}) \ .
\end{equation}
Here again, the second term contributes the same amount as the first.

Therefore, as we have advocated before,
we see that GPDs is naturally related to
the parton distribution of the transverse spin of the proton~\cite{Burkardt:2005hp,Ji:2012sj,Ji:2012vj}.
Recall in the longitudinal case, it is the quark and gluon helicity distributions 
and the twist-three OAM distributions which provide the partonic picture for the proton helicity~\cite{Ji:2020ena}. 

\section{Conclusion}
In this article, we discuss the sum rule
for the transversely-polarized proton. For the intrinsic transverse
AM or spin, we obtain
a similar sum rule as that for the helicity,
except now the quark spin and orbital contributions are frame dependent, with former vanishing
in the infinite-momentum limit. In the IMF, a partonic sum rule can further be derived for
the part of the transverse spin in terms of twist-two quark and gluon
OAM distributions expressible in term of GPDs.

{\it Acknowledgment.}---We thank M. Burkardt and Y. Hatta for discussions related to transverse
spin. This material is supported by the U.S. Department of Energy, Office of Science, Office of Nuclear Physics, under contract number DE-AC02-05CH1123 and DE-SC0020682, and within the framework of the TMD Topical Collaboration.


\begin{thebibliography}{27}
\expandafter\ifx\csname natexlab\endcsname\relax\def\natexlab#1{#1}\fi
\expandafter\ifx\csname bibnamefont\endcsname\relax
  \def\bibnamefont#1{#1}\fi
\expandafter\ifx\csname bibfnamefont\endcsname\relax
  \def\bibfnamefont#1{#1}\fi
\expandafter\ifx\csname citenamefont\endcsname\relax
  \def\citenamefont#1{#1}\fi
\expandafter\ifx\csname url\endcsname\relax
  \def\url#1{\texttt{#1}}\fi
\expandafter\ifx\csname urlprefix\endcsname\relax\def\urlprefix{URL }\fi
\providecommand{\bibinfo}[2]{#2}
\providecommand{\eprint}[2][]{\url{#2}}

\bibitem[{\citenamefont{Jaffe and Manohar}(1990)}]{Jaffe:1989jz}
\bibinfo{author}{\bibfnamefont{R.~L.} \bibnamefont{Jaffe}} \bibnamefont{and}
  \bibinfo{author}{\bibfnamefont{A.}~\bibnamefont{Manohar}},
  \bibinfo{journal}{Nucl. Phys.} \textbf{\bibinfo{volume}{B337}},
  \bibinfo{pages}{509} (\bibinfo{year}{1990}).

\bibitem[{\citenamefont{Ji}(1997{\natexlab{a}})}]{Ji:1996ek}
\bibinfo{author}{\bibfnamefont{X.-D.} \bibnamefont{Ji}},
  \bibinfo{journal}{Phys. Rev. Lett.} \textbf{\bibinfo{volume}{78}},
  \bibinfo{pages}{610} (\bibinfo{year}{1997}{\natexlab{a}}),
  \eprint{hep-ph/9603249}.

\bibitem[{\citenamefont{Filippone and Ji}(2001)}]{Filippone:2001ux}
\bibinfo{author}{\bibfnamefont{B.~W.} \bibnamefont{Filippone}}
  \bibnamefont{and} \bibinfo{author}{\bibfnamefont{X.-D.} \bibnamefont{Ji}},
  \bibinfo{journal}{Adv. Nucl. Phys.} \textbf{\bibinfo{volume}{26}},
  \bibinfo{pages}{1} (\bibinfo{year}{2001}), \eprint{hep-ph/0101224}.

\bibitem[{\citenamefont{Bass}(2005)}]{Bass:2004xa}
\bibinfo{author}{\bibfnamefont{S.~D.} \bibnamefont{Bass}},
  \bibinfo{journal}{Rev. Mod. Phys.} \textbf{\bibinfo{volume}{77}},
  \bibinfo{pages}{1257} (\bibinfo{year}{2005}), \eprint{hep-ph/0411005}.

\bibitem[{\citenamefont{Aidala et~al.}(2013)\citenamefont{Aidala, Bass, Hasch,
  and Mallot}}]{Aidala:2012mv}
\bibinfo{author}{\bibfnamefont{C.~A.} \bibnamefont{Aidala}},
  \bibinfo{author}{\bibfnamefont{S.~D.} \bibnamefont{Bass}},
  \bibinfo{author}{\bibfnamefont{D.}~\bibnamefont{Hasch}}, \bibnamefont{and}
  \bibinfo{author}{\bibfnamefont{G.~K.} \bibnamefont{Mallot}},
  \bibinfo{journal}{Rev. Mod. Phys.} \textbf{\bibinfo{volume}{85}},
  \bibinfo{pages}{655} (\bibinfo{year}{2013}), \eprint{1209.2803}.

\bibitem[{\citenamefont{Leader and Lorcé}(2014)}]{Leader:2013jra}
\bibinfo{author}{\bibfnamefont{E.}~\bibnamefont{Leader}} \bibnamefont{and}
  \bibinfo{author}{\bibfnamefont{C.}~\bibnamefont{Lorcé}},
  \bibinfo{journal}{Phys. Rept.} \textbf{\bibinfo{volume}{541}},
  \bibinfo{pages}{163} (\bibinfo{year}{2014}), \eprint{1309.4235}.

\bibitem[{\citenamefont{Deur et~al.}(2019)\citenamefont{Deur, Brodsky, and
  De~Téramond}}]{Deur:2018roz}
\bibinfo{author}{\bibfnamefont{A.}~\bibnamefont{Deur}},
  \bibinfo{author}{\bibfnamefont{S.~J.} \bibnamefont{Brodsky}},
  \bibnamefont{and} \bibinfo{author}{\bibfnamefont{G.~F.}
  \bibnamefont{De~Téramond}}, \bibinfo{journal}{Rept. Prog. Phys.}
  \textbf{\bibinfo{volume}{82}} (\bibinfo{year}{2019}), \eprint{1807.05250}.

\bibitem[{\citenamefont{Jaffe and Ji}(1991{\natexlab{a}})}]{Jaffe:1990qh}
\bibinfo{author}{\bibfnamefont{R.~L.} \bibnamefont{Jaffe}} \bibnamefont{and}
  \bibinfo{author}{\bibfnamefont{X.-D.} \bibnamefont{Ji}},
  \bibinfo{journal}{Phys. Rev.} \textbf{\bibinfo{volume}{D43}},
  \bibinfo{pages}{724} (\bibinfo{year}{1991}{\natexlab{a}}).

\bibitem[{\citenamefont{Jaffe and Ji}(1991{\natexlab{b}})}]{Jaffe:1991kp}
\bibinfo{author}{\bibfnamefont{R.}~\bibnamefont{Jaffe}} \bibnamefont{and}
  \bibinfo{author}{\bibfnamefont{X.-D.} \bibnamefont{Ji}},
  \bibinfo{journal}{Phys. Rev. Lett.} \textbf{\bibinfo{volume}{67}},
  \bibinfo{pages}{552} (\bibinfo{year}{1991}{\natexlab{b}}).

\bibitem[{\citenamefont{Jaffe and Ji}(1992)}]{Jaffe:1991ra}
\bibinfo{author}{\bibfnamefont{R.~L.} \bibnamefont{Jaffe}} \bibnamefont{and}
  \bibinfo{author}{\bibfnamefont{X.-D.} \bibnamefont{Ji}},
  \bibinfo{journal}{Nucl. Phys.} \textbf{\bibinfo{volume}{B375}},
  \bibinfo{pages}{527} (\bibinfo{year}{1992}).

\bibitem[{\citenamefont{Sivers}(1990)}]{Sivers:1989cc}
\bibinfo{author}{\bibfnamefont{D.~W.} \bibnamefont{Sivers}},
  \bibinfo{journal}{Phys. Rev.} \textbf{\bibinfo{volume}{D41}},
  \bibinfo{pages}{83} (\bibinfo{year}{1990}).

\bibitem[{\citenamefont{Collins}(1993)}]{Collins:1992kk}
\bibinfo{author}{\bibfnamefont{J.~C.} \bibnamefont{Collins}},
  \bibinfo{journal}{Nucl. Phys. B} \textbf{\bibinfo{volume}{396}},
  \bibinfo{pages}{161} (\bibinfo{year}{1993}), \eprint{hep-ph/9208213}.

\bibitem[{\citenamefont{Bakker et~al.}(2004)\citenamefont{Bakker, Leader, and
  Trueman}}]{Bakker:2004ib}
\bibinfo{author}{\bibfnamefont{B.}~\bibnamefont{Bakker}},
  \bibinfo{author}{\bibfnamefont{E.}~\bibnamefont{Leader}}, \bibnamefont{and}
  \bibinfo{author}{\bibfnamefont{T.}~\bibnamefont{Trueman}},
  \bibinfo{journal}{Phys. Rev. D} \textbf{\bibinfo{volume}{70}},
  \bibinfo{pages}{114001} (\bibinfo{year}{2004}), \eprint{hep-ph/0406139}.

\bibitem[{\citenamefont{Burkardt}(2003)}]{Burkardt:2002hr}
\bibinfo{author}{\bibfnamefont{M.}~\bibnamefont{Burkardt}},
  \bibinfo{journal}{Int. J. Mod. Phys. A} \textbf{\bibinfo{volume}{18}},
  \bibinfo{pages}{173} (\bibinfo{year}{2003}), \eprint{hep-ph/0207047}.

\bibitem[{\citenamefont{Burkardt}(2005)}]{Burkardt:2005hp}
\bibinfo{author}{\bibfnamefont{M.}~\bibnamefont{Burkardt}},
  \bibinfo{journal}{Phys. Rev. D} \textbf{\bibinfo{volume}{72}},
  \bibinfo{pages}{094020} (\bibinfo{year}{2005}), \eprint{hep-ph/0505189}.

\bibitem[{\citenamefont{Leader}(2012)}]{Leader:2011cr}
\bibinfo{author}{\bibfnamefont{E.}~\bibnamefont{Leader}},
  \bibinfo{journal}{Phys. Rev. D} \textbf{\bibinfo{volume}{85}},
  \bibinfo{pages}{051501} (\bibinfo{year}{2012}), \eprint{1109.1230}.

\bibitem[{\citenamefont{Ji et~al.}(2012{\natexlab{a}})\citenamefont{Ji, Xiong,
  and Yuan}}]{Ji:2012sj}
\bibinfo{author}{\bibfnamefont{X.}~\bibnamefont{Ji}},
  \bibinfo{author}{\bibfnamefont{X.}~\bibnamefont{Xiong}}, \bibnamefont{and}
  \bibinfo{author}{\bibfnamefont{F.}~\bibnamefont{Yuan}},
  \bibinfo{journal}{Phys. Rev. Lett.} \textbf{\bibinfo{volume}{109}},
  \bibinfo{pages}{152005} (\bibinfo{year}{2012}{\natexlab{a}}),
  \eprint{1202.2843}.

\bibitem[{\citenamefont{Ji et~al.}(2012{\natexlab{b}})\citenamefont{Ji, Xiong,
  and Yuan}}]{Ji:2012vj}
\bibinfo{author}{\bibfnamefont{X.}~\bibnamefont{Ji}},
  \bibinfo{author}{\bibfnamefont{X.}~\bibnamefont{Xiong}}, \bibnamefont{and}
  \bibinfo{author}{\bibfnamefont{F.}~\bibnamefont{Yuan}},
  \bibinfo{journal}{Phys. Lett. B} \textbf{\bibinfo{volume}{717}},
  \bibinfo{pages}{214} (\bibinfo{year}{2012}{\natexlab{b}}),
  \eprint{1209.3246}.

\bibitem[{\citenamefont{Hatta et~al.}(2013)\citenamefont{Hatta, Tanaka, and
  Yoshida}}]{Hatta:2012jm}
\bibinfo{author}{\bibfnamefont{Y.}~\bibnamefont{Hatta}},
  \bibinfo{author}{\bibfnamefont{K.}~\bibnamefont{Tanaka}}, \bibnamefont{and}
  \bibinfo{author}{\bibfnamefont{S.}~\bibnamefont{Yoshida}},
  \bibinfo{journal}{JHEP} \textbf{\bibinfo{volume}{02}}, \bibinfo{pages}{003}
  (\bibinfo{year}{2013}), \eprint{1211.2918}.

\bibitem[{\citenamefont{Harindranath et~al.}(2014)\citenamefont{Harindranath,
  Kundu, and Mukherjee}}]{Harindranath:2013goa}
\bibinfo{author}{\bibfnamefont{A.}~\bibnamefont{Harindranath}},
  \bibinfo{author}{\bibfnamefont{R.}~\bibnamefont{Kundu}}, \bibnamefont{and}
  \bibinfo{author}{\bibfnamefont{A.}~\bibnamefont{Mukherjee}},
  \bibinfo{journal}{Phys. Lett. B} \textbf{\bibinfo{volume}{728}},
  \bibinfo{pages}{63} (\bibinfo{year}{2014}), \eprint{1308.1519}.

\bibitem[{\citenamefont{Ji}(1997{\natexlab{b}})}]{Ji:1996nm}
\bibinfo{author}{\bibfnamefont{X.-D.} \bibnamefont{Ji}},
  \bibinfo{journal}{Phys. Rev.} \textbf{\bibinfo{volume}{D55}},
  \bibinfo{pages}{7114} (\bibinfo{year}{1997}{\natexlab{b}}),
  \eprint{hep-ph/9609381}.

\bibitem[{\citenamefont{Tung}(1985)}]{Tung:1985na}
\bibinfo{author}{\bibfnamefont{W.}~\bibnamefont{Tung}},
  \emph{\bibinfo{title}{{Group Theory In Physics}}} (\bibinfo{year}{1985}).

\bibitem[{\citenamefont{Belinfante}(1940)}]{Belinfante:1940}
\bibinfo{author}{\bibfnamefont{F.~J.} \bibnamefont{Belinfante}},
  \bibinfo{journal}{Physica} \textbf{\bibinfo{volume}{7}}, \bibinfo{pages}{449
  } (\bibinfo{year}{1940}), ISSN \bibinfo{issn}{0031-8914}.

\bibitem[{\citenamefont{Ji}(2013)}]{Ji:2013dva}
\bibinfo{author}{\bibfnamefont{X.}~\bibnamefont{Ji}}, \bibinfo{journal}{Phys.
  Rev. Lett.} \textbf{\bibinfo{volume}{110}}, \bibinfo{pages}{262002}
  (\bibinfo{year}{2013}), \eprint{1305.1539}.

\bibitem[{\citenamefont{Ji}(2014)}]{Ji:2014gla}
\bibinfo{author}{\bibfnamefont{X.}~\bibnamefont{Ji}}, \bibinfo{journal}{Sci.
  China Phys. Mech. Astron.} \textbf{\bibinfo{volume}{57}},
  \bibinfo{pages}{1407} (\bibinfo{year}{2014}), \eprint{1404.6680}.

\bibitem[{\citenamefont{Hoodbhoy et~al.}(1999)\citenamefont{Hoodbhoy, Ji, and
  Lu}}]{Hoodbhoy:1998yb}
\bibinfo{author}{\bibfnamefont{P.}~\bibnamefont{Hoodbhoy}},
  \bibinfo{author}{\bibfnamefont{X.-D.} \bibnamefont{Ji}}, \bibnamefont{and}
  \bibinfo{author}{\bibfnamefont{W.}~\bibnamefont{Lu}}, \bibinfo{journal}{Phys.
  Rev.} \textbf{\bibinfo{volume}{D59}}, \bibinfo{pages}{014013}
  (\bibinfo{year}{1999}), \eprint{hep-ph/9804337}.

\bibitem[{\citenamefont{Ji et~al.}(2020)\citenamefont{Ji, Yuan, and
  Zhao}}]{Ji:2020ena}
\bibinfo{author}{\bibfnamefont{X.}~\bibnamefont{Ji}},
  \bibinfo{author}{\bibfnamefont{F.}~\bibnamefont{Yuan}}, \bibnamefont{and}
  \bibinfo{author}{\bibfnamefont{Y.}~\bibnamefont{Zhao}}
  (\bibinfo{year}{2020}), \eprint{2009.01291}.

\end{thebibliography}

\end{document}